\documentstyle[prl,aps]{revtex}
\begin{document}
\draft
\preprint{gr-qc/9911174}
\title{New Regular Black Hole Solution from Nonlinear Electrodynamics}
\author{Eloy Ay\'on--Beato and Alberto Garc\'{\i}a}
\address{Departamento~de~F\'{\i}sica,~Centro~de~Investigaci%
\'on~y~Estudios~Avanzados~del~IPN\\
Apdo. Postal 14-740, 07000 M\'exico DF, MEXICO}
\maketitle

\begin{abstract}
Using a nonlinear electrodynamics coupled to General Relativity a new
regular {\em exact} black hole solution is found. The nonlinear theory
reduces to the Maxwell one in the weak limit, and the solution corresponds
to a charged black hole for $|q|\leq 2s_{{\rm {c}}}m\approx 1.05\,m$, with
metric, curvature invariants, and electric field regular everywhere.
\end{abstract}

\pacs{04.20.Dw, 04.20.Jb, 04.70.Bw}

In the last years an increasing revival of nonlinear electrodynamic theories
is observed \cite{AlamosB-I}. A nonlinear electrodynamics was first proposed
by Born and Infeld \cite{B-I} at the 30's in order to obtain a
finite--energy electron model; they succeeded in determining an electron of
finite radius. After these first achievements very few investigations were
carried out, as Pleba\'{n}ski mentioned in 1970 at the introduction of his
monograph \cite{Ple68}: {\em If in recent times the interest in NLE cannot
be said to be very popular, it is not due to the fact that one could rise
some serious objections against this theory. It is simply rather difficult
in its mathematical formulation, what causes that it is very unlikely to
derive some concrete results in closed form}. The mayor responsible of the
actual revival, more than 25 years later of the moment of formulation of
quoted statement, is the fact that these nonlinear theories appear as
effective theories at different levels of string/M--theory. For instance,
the Born--Infeld action arises as part of the low energy effective action of
the open superstring theory \cite{FradTsey85,Tseytlin86}. Moreover,
generalizations of the quoted finite--energy solutions (BIons) are very
important in the description of D$p$--branes ---solitons of string theory
described by Dirac--Born--Infeld like actions \cite
{Leigh89,CallanMalda98,Gibbons98}. This kind of nonlinear theories has also
been used as inspiration to build up supersymmetric extensions \cite
{GNSS98,GSS99,BHL99}, and non--Abelian generalizations \cite
{Tseytlin97,Park99} (these last ones are used to generalize the colored
solutions of Einstein--Yang--Mills theory \cite{Tripathy99}). More recently,
in the light of the $AdS$/CFT correspondence (cf.~\cite{Malda99} for
a review), the nonlinear electrodynamics string approach has been used to
obtain solutions describing baryon configurations which are consistent with
confinement \cite{Callan99}.

In other frameworks, various important results have been obtained. For
instance, nonlinear electrodynamics ideas are source of inspiration to
formulate other models, as it is the case of Born--Infeld Skyrmions, where
nonlinear terms are essential in order to obtain stationary solutions \cite
{DMS96,NdOO99}. It has been established that Maxwell and Born--Infeld
theories are singled out among all electromagnetic theories since they bear
both dual invariance \cite{GR95,GR96} and ``good propagations'' (in the
sense that excitations propagate without shocks) \cite{Ple68,DMS99}. It is
also remarkable that all the nonlinear electrodynamics satisfy the zeroth
and first laws of black hole mechanics \cite{Rasheed97}.

In this paper we are interested in another less explored application of
nonlinear electrodynamics: namely its coupling to Einstein theory to obtain
regular black hole solutions. Previous attempts on this direction with
nonlinear electrodynamics either have been totally unsuccessful or only
partially solve the singularity problem; in certain approaches the singular
region is artificially cut off from the whole space--time, and in others
cases the singularity is only weakened \cite{Oliveira94,Soleng95,Palatnik97}%
. We would like to point out that the well--known Born--Infeld Lagrangian is
useless in this topic, since it gives rise to a singular black hole
solution, at least in the spherically symmetric case \cite{Infeld}.

The study on global regularity of black hole solutions is quite important in
order to understand the final state of gravitational collapse of initially
regular configurations, since the Penrose cosmic censorship conjecture
claims that if singularities predicted by General Relativity \cite
{H-E,Senov98} occur, they must be dressed by event horizons 
(cf.~\cite{Wald97,JMagli99} for recent reviews). Previous regular black hole
models has been proposed \cite
{Bardeen68,Ayon93,Borde94,BarrFrolov96,MMPSenovilla96,CaboAyon97}; all of
them have been referred to as ``Bardeen black holes'' \cite{Borde97}, since
Bardeen was the first author producing a surprising regular black hole model 
\cite{Bardeen68}. None of these models is an exact solution to Einstein
equations; there is no known physical sources associated with any of them.
Regular black hole solutions to Einstein equations with physically
reasonable sources were reported in \cite{AyonGarcia98,Magli97}. Other
commonly used approaches to avoid the existence of singularities are based
on the search of more general gravity theories. The best candidate today to
produce singularity--free models, even at the classical level, is
string/M--theory, due to its intrinsic non--locality \cite{Tseytlin95}.

We show in this contribution that it is possible to find regular solutions
within General Relativity simply by coupling to the Einstein equations an
appropriate nonlinear electrodynamics, which in the weak field approximation
becomes the usual Maxwell theory. In what follows we shall exhibit a
singularity--free black hole solution to the quoted about system, which
behaves asymptotically as the Reissner--Nordstr\"{o}m solution.

The dynamics of the theory we are using is governed by the action 
\begin{equation}
{\cal S}=\int dv(\,\frac{1}{16\pi }R-\frac{1}{4\pi }{\cal L}(F)),
\label{eq:action}
\end{equation}
where $R$ is scalar curvature, and ${\cal L}$ is a function of $F\equiv 
\frac{1}{4}F_{\mu \nu }F^{\mu \nu }$, where $F_{\mu \nu }$ is the
electromagnetic strength. We would like to recall that there are more
general Lagrangians depending also on the second invariant, $F_{\mu \nu
}^{\ast }F^{\mu \nu }$, but for the objectives of this work it is enough to
consider only an action as the given in (\ref{eq:action}). Alternatively,
one can describe the considered system using another function, obtained by
means of a Legendre transformation \cite{SGP87} 
\begin{equation}
{\cal H}\equiv 2F{\cal L}_{F}-{\cal L},  \label{eq:Leg}
\end{equation}
where, ${\cal L}_{F}\equiv \partial {\cal L}/\partial F$. Defining $P_{\mu
\nu }\equiv {\cal L}_{F}F_{\mu \nu }$, it can be shown that ${\cal H}$ is a
function of $P\equiv \frac{1}{4}P_{\mu \nu }P^{\mu \nu }=({\cal L}_{F})^{2}F$%
, i.e., $d{\cal H}=({\cal L}_{F})^{-1}d(({\cal L}_{F})^{2}F)={\cal H}%
_{P}dP$, where ${\cal H}_{P}\equiv \partial {\cal H}/\partial P$. Using $%
{\cal H}$, the nonlinear electromagnetic Lagrangian is expressed as ${\cal L}%
=2P{\cal H}_{P}-{\cal H}$, depending now on the anti--symmetric tensor $%
P_{\mu \nu }$. The Einstein--nonlinear--electrodynamics field equations
resulting from action (\ref{eq:action}) are 
\begin{equation}
G_{\mu }^{~\nu }=2({\cal H}_{P}P_{\mu \lambda }P^{\nu \lambda }-\delta _{\mu
}^{~\nu }(2P{\cal H}_{P}-{\cal H})),  \label{eq:Ein}
\end{equation}
\begin{equation}
\nabla _{\mu }P^{\alpha \mu }=0.  \label{eq:Max}
\end{equation}
The ${\cal H}$ function has to satisfy the correspondence to Maxwell theory, 
i.e., ${\cal H}\approx P$ for weak fields ($P\ll 1$). We would like
to point out that in this description the usual electromagnetic strength
tensor is given by 
\begin{equation}
F_{\mu\nu }\equiv {\cal H}_{P}P_{\mu \nu },
\end{equation}
this tensor is the physically relevant quantity, although by using the
auxiliary tensor $P_{\mu\nu}$ many results can be achieved more easily.

The particular non--linear electrodynamics source used to derive our regular
black hole solution is determined by the following function ${\cal H}$: 
\begin{equation}
{\cal H}(P)=P\left( 1-\tanh ^{2}\left( s\sqrt[4]{-2q^{2}P}\right) \right) ,
\label{eq:H}
\end{equation}
where $s$ stands for $s\equiv |q|/2m$; $q$ and $m$ are free parameters which
we anticipate to be associated with charge and mass respectively. Notice
that this last function fulfills the correspondence condition to Maxwell
theory. In order to obtain the desired solution, we consider a static and
spherically symmetric configuration 
\begin{equation}
\text{\boldmath$g$}=-\left( 1-\frac{2M(r)}{r}\right) \text{\boldmath$dt$}%
^{2}+\left( 1-\frac{2M(r)}{r}\right) ^{-1}\text{\boldmath$dr$}^{2}+r^{2}%
\text{\boldmath$d\Omega $}^{2},  \label{eq:spher}
\end{equation}
and assume the following ansatz for the anti--symmetric field 
\begin{equation}
P_{\mu \nu }=2\delta _{\lbrack \mu }^{t}\delta _{\nu ]}^{r}D(r).
\label{eq:stat}
\end{equation}
With these choices, equations (\ref{eq:Max}) are easily integrated, 
\begin{equation}
P_{\mu \nu }=2\delta _{\lbrack \mu }^{t}\delta _{\nu ]}^{r}\frac{q}{r^{2}}%
\quad \longrightarrow \quad P=-\frac{D^{2}}{2}=-\frac{q^{2}}{2r^{4}},
\label{eq:dielec}
\end{equation}
where the integration constant was chosen as $q$, since it plays the role of
the electric charge, which becomes apparent from the evaluation of the
electric field $E=F_{tr}={\cal H}_{P}D$. Using expression (\ref{eq:H}) for $%
{\cal H}$, and (\ref{eq:dielec}), the electromagnetic field strength results
in 
\begin{equation}
F_{\mu \nu }=2\delta _{\lbrack \mu }^{t}\delta _{\nu ]}^{r}E(r),\qquad E(r)=%
\frac{q}{4mr^{3}}\left( 1-\tanh ^{2}(q^{2}/2mr)\right) \left( 4mr-q^{2}\tanh
(q^{2}/2mr)\right) .  \label{eq:E}
\end{equation}
From these expressions two conclusions follow: the electric field is regular
everywhere, and asymptotically behaves as $E=q/r^{2}+O(1/r^{4})$, i.e., 
a Coulomb field with electric charge $q$.

The $_{t}^{~t}$ component of Einstein equations (\ref{eq:Ein}) yields 
\begin{equation}
M^{\prime }(r)=-r^{2}{\cal H}(P).  \label{eq:tt}
\end{equation}
Substituting ${\cal H}$ from (\ref{eq:H}), with $P=-q^{2}/2r^{4}$, the first
integral of (\ref{eq:tt}) is 
\begin{equation}
M(r)= - m\tanh (q^{2}/2mr) + K,  \label{eq:int}
\end{equation}
where $K$ is an integration constant which can be evaluated using the mass
definition, $m\equiv \lim_{r\rightarrow \infty }M(r)$, thus $K=m$ and
consequently $M(r)=m\left( 1-\tanh (q^{2}/2mr)\right)$. Entering (\ref
{eq:int}) into (\ref{eq:spher}), one finally obtains the following metric 
\begin{equation}
\mbox{\boldmath$g$}=-\left( 1-\frac{2m\left( 1-\tanh (q^{2}/2mr)\right) }{r}%
\right) \mbox{\boldmath$dt$}^{2}+\left( 1-\frac{2m\left( 1-\tanh
(q^{2}/2mr)\right) }{r}\right) ^{-1}\mbox{\boldmath$dr$}^{2}+r^{2}%
\mbox{\boldmath$d\Omega $}^{2}.  \label{eq:regbh}
\end{equation}
It can be noted that the metric asymptotically behaves as the
Reissner--Nordstr\"{o}m solution, i.e., $%
g_{tt}=1-2m/r+q^{2}/r^{2}+O(1/r^{4})$, thus the parameters $m$ and $q$ can
be correctly associated with mass and charge respectively.

We shall show that for a certain range of the charge parameter our metric
solution (\ref{eq:regbh}) describes a black hole, which in addition is
regular everywhere. Making the substitutions $y=2mr/q^{2}$, $s=|q|/2m$ we
rewrite $g_{tt}$ as 
\begin{equation}
-g_{tt}=A(y,s)\equiv 1-\frac{1-\tanh (1/y)}{s^{2}y}.  \label{eq:A}
\end{equation}
Independently of the nonvanishing value of $s$, the last function has a
single minimum for $y_{{\rm {m}}}\approx 1.56$, which is the only positive
solution for $y$ of the equation $\partial _{y}A(y,s)=0$, which explicitly
amounts to 
\[
\left( 1-\tanh (1/y)\right) \left( 1-y+\tanh (1/y)\right) =0. 
\]
Next, the solution of the equation $A(y_{{\rm {m}}},s)=0$ gives the single
positive root $s_{{\rm {c}}}\approx 0.53$. At $y_{{\rm {m}}}$, for $s<s_{%
{\rm {c}}}$ the quoted minimum is negative, for $s=s_{{\rm {c}}}$ the
minimum vanishes and for $s>s_{{\rm {c}}}$ the minimum is positive. From the
analytical expressions of the curvature invariants for metric (\ref{eq:regbh}%
) 
\begin{equation}
R=\frac{2\tanh (1/y)\left( 1-\tanh ^{2}(1/y)\right) }{q^{2}s^{4}y^{5}},
\label{eq:R}
\end{equation}
\begin{equation}
R_{\mu \nu }R^{\mu \nu }=\frac{2\left( 1-\tanh ^{2}(1/y)\right) ^{2}}{%
q^{4}s^{8}y^{10}}\left( 2y^{2}-2y\tanh (1/y)+\tanh ^{2}(1/y)\right) ,
\label{eq:Ricci}
\end{equation}
\begin{eqnarray}
R_{\mu \nu \alpha \beta }R^{\mu \nu \alpha \beta }=\frac{4\left( 1-\tanh
(1/y)\right) ^{2}}{q^{4}s^{8}y^{10}}\left( \!\frac{{}}{{}}\right. \!
&&\!3y^{4}+y^{2}\left( \tanh (1/y)+1\right) \left( 7\tanh (1/y)+5-6y\right) +
\nonumber \\
&&\ \ \ \left. \tanh (1/y)\left( \tanh (1/y)+1\right) ^{2}\left( \tanh
(1/y)-4y\right) \right) ,  \label{eq:Riemman}
\end{eqnarray}
one concludes they are all regular everywhere. Hence, for $s{\leq}s_{{\rm {c}%
}}$ the singularities appearing in (\ref{eq:regbh}) due to the vanishing of $%
A$ are only coordinate--singularities describing the existence of event
horizons, consequently, we are in the presence of black hole solutions for $%
|q|\leq 2s_{{\rm {c}}}m\approx 1.05\,m$. It should be notice that metrics
possessing regular standard invariants, as (\ref{eq:R})--(\ref{eq:Riemman}),
could still present non--regular behavior of the differential invariants, $%
R_{;\alpha_{1}\cdots\alpha_{n}}R^{;\alpha _{1}\cdots \alpha _{n}}$, $R_{\mu
\nu ;\alpha _{1}\cdots \alpha _{n}}R^{\mu \nu ;\alpha _{1}\cdots \alpha
_{n}} $, $R_{\mu \nu \alpha \beta ;\alpha _{1}\cdots \alpha _{n}}R^{\mu \nu
\alpha \beta ;\alpha _{1}\cdots \alpha _{n}}$ \cite{MusgLake95}. Because of
the structure of the metric components of our solution (\ref{eq:regbh}) and
their derivatives, the components of the Riemman tensor behave generically
at $r=0$ as $\sim g(r)/r^{a}$, where $g(r)$ is a function vanishing
exponentially as $r\rightarrow 0$. This fact explain the bounded behavior of
the invariants (\ref{eq:R})--(\ref{eq:Riemman}). Moreover, the quoted
behavior at $r=0$ extends to the covariant differential of any order of the
Riemman tensor (the derivative of an exponentially vanishing function is
again an exponentially vanishing function, and the contributions due to the
Christoffel symbols entering in the covariant derivatives do not alter the
quoted pattern). In its turn, this again yields a bounded behavior on the
differential invariants of our solution.

For the corresponding values of mass and charge we have, under the strict
inequality $|q|<2s_{{\rm {c}}}m$, inner and event horizons for the Killing
field $\text{\boldmath$k$}=\text{\boldmath$\partial /\partial {t}$}$,
defined by $k_{\mu }k^{\mu }=g_{tt}=0$. For the equality $|q|=2s_{{\rm {c}}%
}m $, the horizons shrink into a single one, where also $\nabla _{\nu
}(k_{\mu }k^{\mu })=0$, i.e., this case corresponds to an extreme
black hole as in the Reissner--Nordstr\"{o}m solution. The extension of the
metric beyond the horizons, up to $r=0$, becomes apparent by passing to the
standard advanced and retarded Eddington--Finkelstein coordinates, in terms
of which the metric is well--behaved everywhere, even in the extreme case.
The maximal extension of this metric can be achieved by following the main
lines presented in \cite{Chandra83} for the Reissner--Nordstr\"{o}m
solution, taking care of course, of the more involved integration in our
case of the tortoise coordinate $r^{\ast }\equiv \int A^{-1}dr$. The global
structure of our space--time is similar to the structure of the
Reissner--Nordstr\"{o}m black hole, except that the usual singularity of
this solution, at $r=0$, has been smoothed out and now it simply corresponds
to the origin of the spherical coordinates.

It must be noted that for $q=0$ our solution (\ref{eq:regbh}) becomes the
singular Schwarzschild one. For $q\neq 0$, in the derived solution, the
charge appears by means of a nonlinear electromagnetic field $F_{\mu \nu }$,
which is singularity--free and regularizes the associated gravitational
field. In nonlinear electrodynamics, one can think of the space--time
coupled to the nonlinear field as produced by a charge distribution through
the whole space; in each point of it, the nonlinear interaction of
gravitation and electromagnetism is responsible for the regular character of
the fields involved. One arrives at this effective charge distribution from
the non--zero divergence of the electric field, as can be established from
the Maxwell equations (\ref{eq:Max}). Nonlinear electrodynamics can be
entirely constructed in terms of functions of $F_{\mu \nu }$ and its
invariants. Historically the introduction of the auxiliary field $P_{\mu \nu
}$ obeyed the necessity of establishing a relationship between the new
electromagnetic theory with theories currently at hand, as the Maxwell
theory of continuous media, through the so--called ``material equations'' 
\cite{B-I,Ple68}. In turn this $P_{\mu \nu }$ field resulted to be useful in
the derivation of exact solutions in General Relativity \cite{SGP87};
Maxwell equations (\ref{eq:Max}) are linear in terms of $P_{\mu \nu }$.
Since $P_{\mu \nu }$ is a secondary field ---function of $F_{\mu \nu }$---
the presence of a Coulomb--like singularity in it is irrelevant, the physics
has to be extracted from $F_{\mu \nu }$. Since $F_{\mu \nu }$ is regular
everywhere the conclusion we arrive at is that the nonlinearity of the
electrodynamic source together with the well--known nonlinearity of the
gravitational field are the only responsible for the regular behavior of the
metric and the matter content. This last conclusion is very important, since
indicates that singularities occurring in Physics can be due to
extrapolations of the linear interactions to strong regimes, where a
nonlinear description would be more appropriate.

\acknowledgments
This work was partially supported by the CONACyT Grant ``Black Holes,
Inflation and Matter,'' and a fellowship from the Sistema Nacional de
Investigadores (SNI). E.A.B. thanks the staff of the Physics Department at
CINVESTAV for support.

\end{document}